\documentclass[amsmath,amssymb,aps,prb,floatfix,superscriptaddress]{revtex4-1}

\usepackage{graphicx}
\usepackage{stmaryrd}
\usepackage{dcolumn}
\usepackage{bm}
\usepackage[usenames,dvipsnames]{xcolor}
\usepackage{multirow}
\usepackage{times}
\usepackage{chngcntr}
\usepackage{hyperref}
\usepackage{lineno} 
\usepackage{url,hyperref}
\usepackage{braket}

\newcommand{\code}[1]{\texttt{#1}}

\begin{document}

\title[The AiiDA-Spirit plugin]{The AiiDA-Spirit plugin for automated spin-dynamics simulations and multi-scale modelling based on first-principles calculations} 

\author{Philipp Rüßmann}
\email[Corresponding author: ]{p.ruessmann@fz-juelich.de}
\affiliation{Peter Grünberg Institut (PGI-1) and Institute for Advanced Simulation (IAS-1), 
	Forschungszentrum Jülich and JARA, D-52425 Jülich, Germany}
\author{Jordi Ribas Sobreviela}
\affiliation{Peter Grünberg Institut (PGI-1) and Institute for Advanced Simulation (IAS-1), 
	Forschungszentrum Jülich and JARA, D-52425 Jülich, Germany}
\affiliation{RWTH Aachen University, Aachen, Germany}
\author{Moritz Sallermann}
\affiliation{Peter Grünberg Institut (PGI-1) and Institute for Advanced Simulation (IAS-1), 
	Forschungszentrum Jülich and JARA, D-52425 Jülich, Germany}
\affiliation{Science Institute and Faculty of Physical Sciences, University of Iceland, VR-III, 107 Reykjavík, Iceland}
\author{Markus Hoffmann}
\affiliation{Peter Grünberg Institut (PGI-1) and Institute for Advanced Simulation (IAS-1), 
	Forschungszentrum Jülich and JARA, D-52425 Jülich, Germany}
\author{Florian Rhiem}
\affiliation{Peter Grünberg Institut / Jülich Centre for Neutron Science - Technical Services and Administration (PGI/JCNS-TA), 
	Forschungszentrum Jülich and JARA, D-52425 Jülich, Germany}
\author{Stefan Blügel}
\affiliation{Peter Grünberg Institut (PGI-1) and Institute for Advanced Simulation (IAS-1), 
	Forschungszentrum Jülich and JARA, D-52425 Jülich, Germany}

\begin{abstract}

Landau-Lifshitz-Gilbert (LLG) spin-dynamics calculations based on the extended Heisenberg Hamiltonian is an important tool in computational materials science involving magnetic materials. LLG simulations allow to bridge the gap from expensive quantum mechanical calculations with small unit cells to large supercells where the collective behavior of millions of spins can be studied.
In this work we present the AiiDA-Spirit plugin that connects the spin-dynamics code \code{Spirit} to the AiiDA framework. AiiDA provides a Python interface that facilitates performing high-throughput calculations while automatically augmenting the calculations with metadata describing the data provenance between calculations in a directed acyclic graph. The AiiDA-Spirit interface thus provides an easy way for high-throughput spin-dynamics calculations. The interface to the AiiDA infrastructure furthermore has the advantage that input parameters for the extended Heisenberg model can be extracted from high-throughput first-principles calculations including a proper treatment of the data provenance that ensures reproducibility of the calculation results in accordance to the FAIR principles.
We describe the layout of the AiiDA-Spirit plugin and demonstrate its capabilities using selected examples for LLG spin-dynamics and Monte Carlo calculations. Furthermore, the integration with first-principles calculations through AiiDA is demonstrated at the example of $\gamma$--Fe, where the complex spin-spiral ground state is investigated.

\end{abstract}

\maketitle

\section{Introduction}

Magnetic materials play an important role in modern technology. Their most important applications range from electrical motors to the storing and processing of digital information. The performance of such applications crucially relies on the performance of magnets where the knowledge of their magnetic order, the Curie temperature, the magnetic hardness or their chirality plays an important role. 
Computational materials design of magnetic materials and devices is a complex multi-scale problem. While quantum mechanical calculations allow to predict the interaction strength among magnetic atoms \cite{Liechtenstein1987}, large scale simulations for nanometer to micrometer length scales are unfeasible due to their computational cost. Mapping these interactions to a Heisenberg model allows to bridge the scales from the atomic length scale to the length scale of devices.

Spin-dynamics calculations based on the Landau-Lifshitz-Gilbert (LLG) equation \cite{LandauLifshitz, Gilbert} are a widely used tool for this multi-scale modelling of magnetic materials \cite{dupe2014tailoring, hoffmann2017antiskyrmions, hoffmann2021racetrack, weissenhofer2021skyrmion},
providing access to the collective behavior of millions of spins \cite{spirit-paper}. This approach allows to find, for instance, the (non-collinear) magnetic ground state based on an energy minimization of the extended Heisenberg Hamiltonian or to study the dynamics of magnetic solitons such as skyrmions~\cite{Muhlbaue2009skyrmion,tokura2010realspace,heinze2011spontaneous,back2020roadmap} or hopfions~\cite{bogolubsky1988three,sutcliffe2018hopfions,kent2021creation} at finite temperature.
In combination with the geodesic nudged elastic band method and the harmonic transition state theory \cite{bessarab2015gneb,
bessarab2012htst}
it furthermore gives insight into the stability of aforementioned objects \cite{spirit-paper}.

In this work we introduce the AiiDA-Spirit plugin that connects the \code{Spirit} code \cite{spirit-code} to the AiiDA environment \cite{aiida1}. AiiDA is an open-source Python framework designed around the FAIR principles of findable, accessible, interoperable and reusable data \cite{Wilkinson2016} in computational science \cite{aiida0}. Calculations that run through the AiiDA infrastructure are automatically stored as nodes in a database together with all inputs and outputs that are necessary to reproduce the simulation results. This results in an directed acyclic graph that can connect different nodes which can be used to reproduce the data provenance from a final result.

In the context of spin-dynamics simulations, a simulation result could be the magnetic ordering obtained from a minimization of the forces on each spin in an LLG calculation. The outcome of such a simulation will in general depend on input parameters such as the geometry (positions of the spins, size of simulation cell, open or periodic boundary conditions), the exchange coupling constants, or applied external fields as well as temperature noise. But also the starting point for the minimization (e.g.\ starting from an ordered ferromagnet or from random spin orientations) are important as local minima in the energy landscape can generally be present in which metastable states can be stabilized. To ensure reproducible calculation results,  keeping track of the full data provenance of a simulation is necessary.

AiiDA's plugin infrastructure allows to orchestrate and combine different sequences of calculations, possibly using different simulation software and methods, through a common interface. Here, we use this to first generate exchange coupling parameters from DFT calculations using the \code{JuKKR} code \cite{jukkr} with the help of the AiiDA-KKR plugin~\cite{aiida-kkr-paper, aiida-kkr-code}. Then, we proceed with spin-dynamics simulations using the \code{Spirit} code \cite{spirit-paper, spirit-code} via the newly developed AiiDA-Spirit plugin~\cite{aiida-spirit}. This allows to include the full history of the input parameter generation for spin-dynamics calculations in the provenance graph of a Spirit simulation. Using AiiDA therefore facilitates multi-scale modelling that combines the predictive power of DFT calculations and the speed and scalability of spin-dynamics simulations in the same framework.

The AiiDA engine \cite{aiida2} provides a highly scalable infrastructure that is able to deal with thousands of calculations simultaneously. Together with the simple Python interface that AiiDA-Spirit provides, spin-dynamics simulations are possible in an automated way which can be used in a high-throughput fashion. This opens new possibilities for applying the \code{Spirit} code in automated setups and as part of complex workflows in conjunction with other simulation methods such as DFT. This new capability allows to integrate \code{Spirit} in the toolbox of methods that are used in automated computational materials design for magnetic materials \cite{Himanen2019}. 

This paper is structured as follows. First the methods section introduces the theory behind spin-dynamics simulations. Then the AiiDA-Spirit plugin is presented which is then applied to (i) a parameter exploration based on a toy model and a large number of high-throughput AiiDA-Spirit calculations, (ii) to a simple Monte Carlo example to find the critical temperature of a model system, and (iii) multi-scale modelling combining DFT and LLG calculations at the example of $\gamma$--Fe. Finally the paper concludes with a discussion of the results.

\section{Methods}

\subsection{Spirit theory}

All spin-dynamics simulations shown throughout the paper were performed with the \code{Spirit} code \cite{spirit-code,spirit-paper}. The \code{Spirit} code provides a framework for atomic-scale spin simulations and combines both a graphical user interface as well as an easy accessible Python API. All simulations performed with Spirit are based on an extended Heisenberg Hamiltonian describing the interaction of spins $\vec{S}_{i}=\vec{M}_i/\mu_{i}$ ($\mu_i=\lvert\vec{M}_i\rvert$) sitting at lattice sites $i$. It can be written in its most general form as
\begin{eqnarray}
    H = &-&\sum_{\braket{ij}}J_{ij} \left( \vec{S}_i \cdot \vec{S}_j \right) -\sum_{\braket{ij}}\vec{D}_{ij} \cdot \left( \vec{S}_i \times \vec{S}_j \right)\nonumber \\
    &-& \sum_{i} K_{\perp} (\vec{S}_i \cdot \hat{\vec{e}}_z)^2 - \sum_{i} \mu_{i} \vec{B}\cdot \vec{S}_{i} \nonumber \\
    &-& \frac{\mu_{0}}{8\pi}\sum_{i\neq j} \frac{3\left(\vec{S}_i \cdot \hat{\vec{r}}_{ij}\right)\left(\vec{S}_j \cdot \hat{\vec{r}}_{ij}\right)-\vec{S}_i\cdot\vec{S}_j}{r_{ij}^3} \nonumber \\
    &-& \sum_{\braket{ijkl}} K_{ijkl} \left(\vec{S}_i \cdot \vec{S}_j\right)\left(\vec{S}_k \cdot \vec{S}_l\right)\quad .
    \label{eq:Heisenberg}
\end{eqnarray} 
Here, the first line contains the isotropic and anisotropic exchange interactions, the later also referred to as Dzyaloshinskii-Moriya interaction. The second and third line describe the on-site anisotropy, the Zeeman energy due to an external magnetic field $\vec{B}$, as well as the dipolar contribution. The last term allows to include higher-order exchange interactions~\cite{hoffmann2020systematic} such as the conventional four-spin or four-spin-four-site interaction~\cite{heinze2011spontaneous}, the four-spin-three-site interaction~\cite{kronlein2018magnetic}, as well as the biquadratic interaction~\cite{szilva2013interatomic}. The list of pairs $\braket{ij}$ and quadruplets $\braket{ijkl}$ as well as their respective parameters $J_{ij}$, $\vec{D}_{ij}$, and $K_{ijkl}$ can be defined by the user based on the desired use case. Furthermore, the system geometry such as the lattice symmetry and lattice size can be chosen arbitrarily and \code{Spirit} allows to introduce defects such as vacancies or atoms of different types.
To obtain ground state as well as thermal properties of the investigated system, either the Monte Carlo method based on a Metropolis algorithm or Landau-Lifshitz-Gilbert dynamics can be used.

A more detailed description of the Spirit framework as well as its further functionalities, such as the possibility to calculate lifetimes of magnetic textures based on the combination of geodesic nudged elastic band and harmonic transition state theory calculations, can be found in Ref.~\cite{spirit-paper}.

\subsection{The AiiDA-Spirit plugin}

AiiDA's plugin system allows to combine various simulation codes and methods (to date more than 60 plugins exist already \cite{plugin-registry}) on the same footing while augmenting the calculation done through the AiiDA infrastructure with the stored data provenance. 
Albeit their significance in research on magnetic materials, spin-dynamics calculations have not been at the center of the AiiDA community so far. To the best of our knowledge, apart from the AiiDA-Spirit plugin we introduce here only an initial version of the AiiDA-UppASD plugin \cite{uppasd-aiida} to the \code{UppASD} code \cite{Skubic_2008} exists which is able to connect AiiDA to a spin-dynamics simulation engine.

\begin{figure}[htb]
    \centering
    \includegraphics[width=0.8\linewidth]{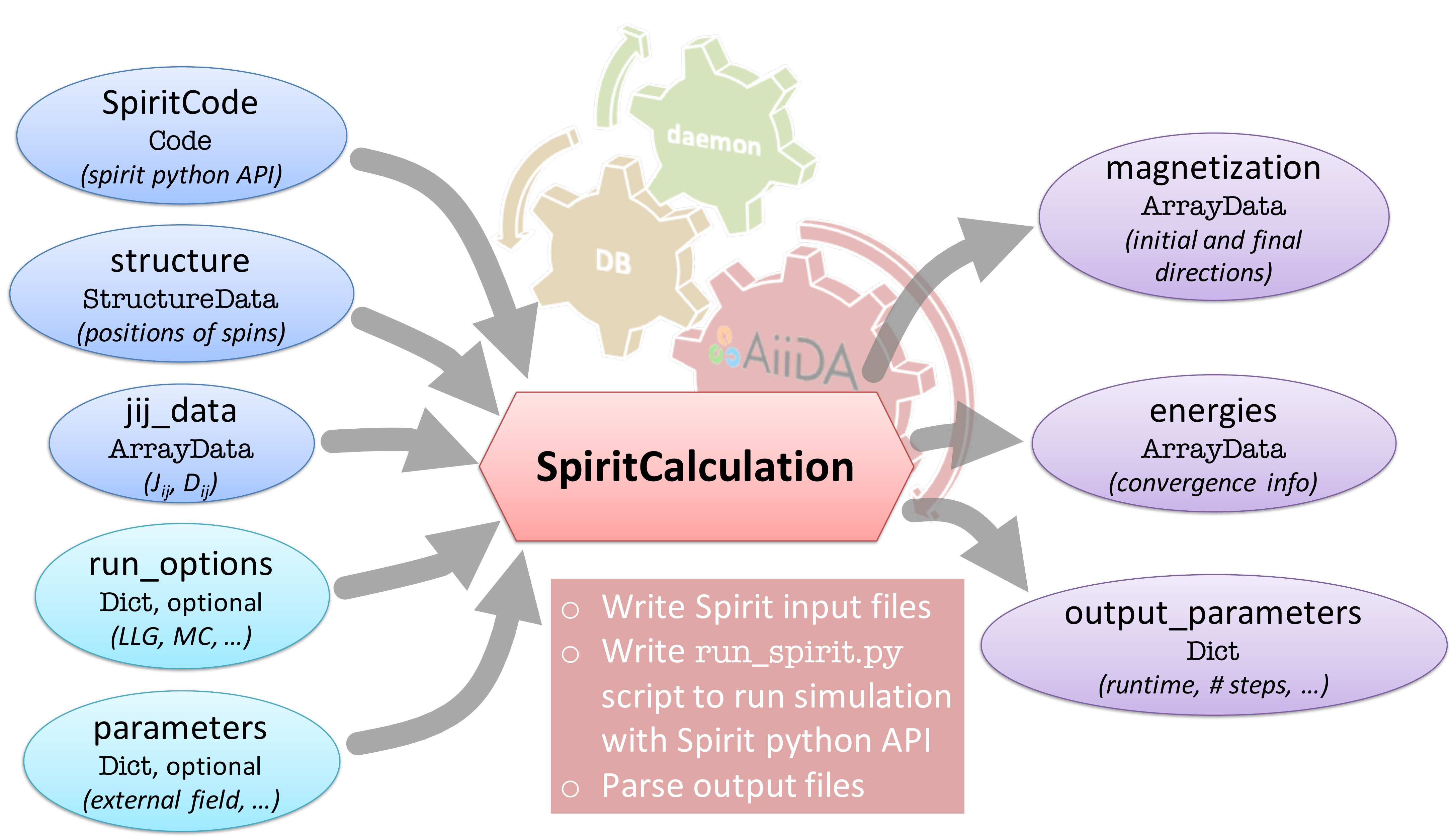}
    \caption{Layout of the {SpiritCalculation} that is at the heart of the AiiDA-Spirit plugin. On the left hand side the possible input nodes are shown which are translated by the {SpiritCalculation} into the appropriate input files needed to execute Spirit. The {run\_options} and {parameters} input nodes are optional that default to a basic LLG calculation starting from random orientation of the spins and without external fields or temperature. The typical output nodes for a LLG calculation are shown on the right hand side.}
    \label{fig:SpiritCalc}
\end{figure}

In the context of AiiDA, a calculation plugin needs to be able to generate typical input files that are required to run a calculation through a bash script that will be generated when a calculation is submitted to a computer or as a job on a supercomputer. At the heart of the AiiDA-Spirit plugin lies the \code{SpiritCalculation} that connects the Spirit code via the Spirit Python API to AiiDA. The Layout of the \code{SpiritCalculation} is shown in Fig.~\ref{fig:SpiritCalc}. To run a Spirit calculation a number of input nodes are required:
\begin{itemize} \setlength\itemsep{0em}
    \item a \code{structure} node describing the lattice of spins (i.e.\ their positions in the unit cell), 
    \item an array of the corresponding \code{jij\_data} that contains the pairwise $J_{ij}$ and $\vec{D}_{ij}$ parameters for the extended Heisenberg Hamiltonian (\ref{eq:Heisenberg}), 
    \item the \code{SpiritCode} that is an installation of the Spirit Python API on the computer where the calculation should run, 
    \item and \code{run\_options} as well as input \code{parameters} that control the type of the Spirit run (e.g.\ LLG or Monte Carlo) or further settings like strength and direction of external fields, respectively.
\end{itemize}
Additionally, input modes that trigger special features of the Spirit code such as disorder and defects in the structure or pinning of spins to certain directions can be controlled with the corresponding optional input nodes. The \code{SpiritCalculation} then implements the functionality to translate this information into the appropriate input files and runs the calculation using the Spirit Python API. The AiiDA daemon automatically takes care of creating a suitable job script, copying necessary input files, and of submitting and monitoring the calculation run. Once the calculation job finishes, important output files are copied back to the {retrieved} folder in the AiiDA file repository associated to the AiiDA database. Then, the \code{SpiritParser} extracts useful information that should be stored in the database. For the example of a LLG calculation this entails settings such as the number of LLG steps until convergence, the used wall clock time on the computer where the calcualtion ran, an array of the \code{energies} (i.e.\ exchange energy per spin), and the initial and final directions of the spins in the \code{magnetization} array. 

Apart from the \code{SpiritCalculation} and \code{SpiritParser}, AiiDA-Spirit comes with some tools that can be used in the typical jupyter notebook environment that is often used in the context of AiiDA. In particular we mention the \code{show\_spins} tool of AiiDA-Spirit which provides the Spirit visualization capabilities in a simple Python API. This consists of a WebAssembly and WebGL version of the VFRendering library \cite{vfrendering} in combination with a JavaScript interface that can be used to visualize the directions of the spins from the web-browser based environment natural to jupyter notebooks.

\subsection{DFT-based calculation of exchange coupling constants}

The density functional theory (DFT) results of this work were produced within the generalized gradient approximation (GGA-PBE) \cite{PBE} using the full-potential scalar-relativistic Korringa-Kohn-Rostoker Green's function method (KKR) \cite{Ebert2011} as implemented in the \code{JuKKR} code package \cite{jukkr}. We use an $\ell_{max} = 3$ cutoff in the angular momentum expansion with an exact description of the atomic cells \cite{Stefanou1990,Stefanou1991}.
After the self-consistent DFT calculations, the method of infinitesimal rotations \cite{Liechtenstein1987} was used to compute the exchange interaction parameters $J_{ij}$.
The series of DFT calculations in this study are orchestrated using the {AiiDA-KKR} \cite{aiida-kkr-paper} plugins to the AiiDA infrastructure \cite{aiida1}. The complete dataset that includes the full provenance of the calculations is made publicly available in the materials cloud repository \cite{Talirz2020, doi-dataset}.

\section{Results}


\subsection{Automated LLG calculations for model parameter exploration}

To illustrate the usage of AiiDA-Spirit we first consider a toy model consisting of a single layer of spins in a simple-cubic lattice. The complete example is part of the dataset that accompanies this publication \cite{doi-dataset}. We assume only nearest neighbor interactions with isotropic exchange $J_{1}=10\,\mathrm{meV}$ and Dzyaloshinskii-Moriya interactions with a strength of $D_{1}=6\,\mathrm{meV}$. This choice of parameters does not reflect any concrete physical system but is chosen for illustration purposes because it is known to produce skyrmions with small radii. The generation of the corresponding input node for the \code{SpiritCalculation} where including the directions of the DMI vectors can be seen in the following code snippet.
\begin{small}
\begin{verbatim}
    jijs_expanded = np.array([
        #i  j da db dc  Jij   Dx   Dy   Dz
        [0, 0, 1, 0, 0, 10.0, 6.0, 0.0, 0.0],
        [0, 0, 0, 1, 0, 10.0, 0.0, 6.0, 0.0],
    ])
    jij_data = aiida.orm.ArrayData()
    jij_data.set_array('Jij_expanded', jijs_expanded)
\end{verbatim}
\end{small}

For this example we consider an isolated layer of spins with periodic boundary conditions in the plane. We choose a supercell for the \code{SpiritCalculation} of $50\times50\times1$ spins. Furthermore we apply an external field of various strength (in the code snippet we show the input parameters for a value of $25\,\mathrm{T}$) in the direction perpendicular to the film in the following code snippet.
\begin{small}
\begin{verbatim}
    parameters = aiida.orm.Dict(
        dict={
            'n_basis_cells': [50, 50, 1],
            'boundary_conditions': [True, True, False],
            'external_field_magnitude': 25.0,
            'external_field_normal': [0.0, 0.0, 1.0],
            'mu_s': [2.0], # one value per spin in the unit cell
        })
\end{verbatim}
\end{small}

Starting from random orientations of the spins we then perform a time evolution using the LLG method with the Depondt solver~\cite{depondt2009spin}. The parameters for the LLG calculations are summarized in Table~\ref{tab:LLGparam}.

\begin{table}[htb]
\centering
\caption{Parameters for the LLG calculations of the toy model. Arrays are indicated by the square brackets. Except for    \code{external\_field\_magnitude} and \code{llg\_temperature} all parameters are kept fixed in the simulations.}
\begin{tabular}{c|c|c}
    Parameter & Value & Description \\\hline
    \code{n\_basis\_cells} & $[50, 50, 1]$ & Size of the simulation cell \\
    \code{boundary\_conditions} & [True, True, False] & Periodic boundary conditions \\
    \code{llg\_n\_iterations}       & $100000$ & Number of iterations \\
    \code{llg\_damping}             &  $0.3$   &  Damping constant \\
    \code{llg\_beta}                &  $0.1$   &  Non-adiabatic damping \\
    \code{llg\_dt}                  &  $0.001$ & Time step dt [ps] \\
    \code{llg\_force\_convergence}  &  $10^{-7}$ & Force convergence parameter \\
    \code{llg\_temperature}         & $0 \dots 75 $  &  Temperature [K] \\
    \code{external\_field\_magnitude} & $-50 \dots 50$ &  Magnitude of the external field \\
    \code{external\_field\_normal} & $[0.0, 0.0, 1.0]$ & Direction of the external field \\
    \code{mu\_s} & $[2.0]$ & Spin moment [$\mu_B$]
\end{tabular}
\label{tab:LLGparam}
\end{table}

To harness the high-throughput capabilities of the  AiiDA-Spirit plugin we perform a series of \code{SpiritCalculations} to screen a range of external fields and temperatures. We change the temperature from $0$ to $75\,\mathrm{K}$ in $2.5\,\mathrm{K}$ steps and vary the external field from $-50\,\mathrm{T}$ to $+50\,\mathrm{T}$ in steps of $2.5\,\mathrm{T}$. The calculations for each parameters set are repeated 5 times starting from different random orientations of the spins for statistical averaging. This amounts to $31 \times 41 \times 5 =6355$ individual \code{SpiritCalculations} that were submitted to a in-house compute cluster. We stress that the AiiDA daemon \cite{aiida2} conveniently takes care of creating submission scripts and automatically retrieves and parses the outcome of the calculations without the need for any user interaction. 

In order to analyze the outcome of the \code{SpiritCalculations} we chose to investigate the topological charge in the simulation cell at the end of the LLG simulation. For a continuous vector field $\vec{m}$ it is defined as
\begin{equation}
    \rho_\mathrm{T} = \frac{1}{4\pi}\int \vec{m} \cdot \left( \frac{\partial\vec{m}}{\partial x} \times \frac{\partial\vec{m}}{\partial y} \right) \mathrm{d}x\mathrm{d}y\,\,.
    \label{eq:topocharge}
\end{equation}
We added a custom post-processing step to the \code{SpiritCalculation} which uses the \code{get\_topological\_charge} function of the spirit Python API. This function calculates the topological charge from the discretized form of (\ref{eq:topocharge}) as a summation over all contributions of triangles formed by neighboring spins in the simulation cell \cite{spirit-paper}.

\begin{figure}[htb]
    \centering
    \includegraphics[width=1.0\linewidth]{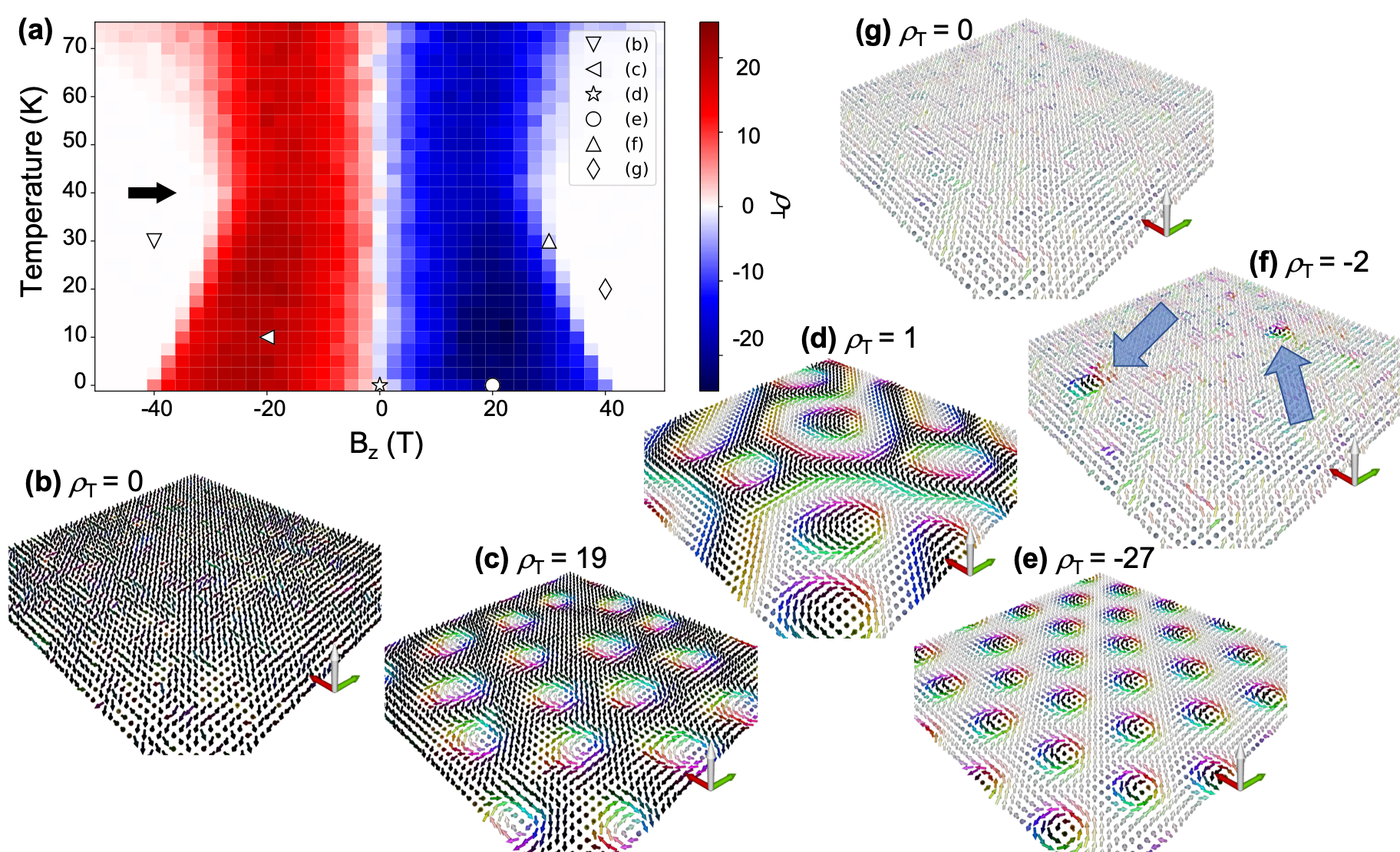}
    \caption{Topological charge of the toy model discussed in the text. (a) Dependence of the topological charge $\rho_{\mathrm{T}}$ on the external magnetic field and temperature calculated from the final spin texture after an LLG calculation. The black arrow highlights the inflection point where $|\rho_\mathrm{T}(B_z)|$ has a minimum with respect to $B_z$. For the parameters marked by the symbols (b-g) the resulting spin textures are shown in the corresponding panels. The arrows in (f) highlight the two skyrmions that result in a topological charge of $\rho_{\mathrm{T}}=-2$.
    }
    \label{fig:example}
\end{figure}

Figure~\ref{fig:example} shows the outcome of these simulations where the topological charge is shown for all 1271 pairs $(T, B_z)$ together with selected spin configurations of representative calculations marked by the symbols (b)--(g). The real-space spin configuration at the end of the LLG calculations were visualized using the \code{show\_spins} tool of the AiiDA-Spirit plugin. It can be seen that a small external field leads to the appearance of skyrmions which in this case have a topological charge of $\pm 1$, depending whether they form in a ferromagnetic background of spins pointing in $-z$ (c) or $+z$ (e) direction.
In general, the topological charge counts the difference between the amount of skyrmions in up-domains  ($\rho_\mathrm{T}>0$) and skyrmions in down-domains  ($\rho_\mathrm{T}<0$) as seen for vanishing external field in (d) where several skyrmions with opposite topological charges lead to a near cancellation of the total topological charge.

At very large fields, the Zeeman exchange coupling term becomes larger than the DMI energy and a homogeneous ferromagnet forms (b, g). Temperature fluctuations tend to deform the skyrmions (f) and destabilize them. Thus, at elevated temperatures smaller magnitudes of the external magnetic field leads to a vanishing topological charge. As highlighted by the black arrow in (a), this is however only true up to a certain critical temperature. For $T>40\,\mathrm{K}$ the topological charge increases again which can be explained by the energy barrier of skyrmion formation and destruction. At these elevated temperatures the fluctuations of the spin directions are larger than $40\,\mathrm{K} \cdot  k_\mathrm{B} \approx 3.45\,\mathrm{meV}$ which we conjecture is the energy barrier for skyrmion formation. While the energy barrier can in principle be calculated by performing geodesic nudged elastic band calculations, this is beyond the scope of this paper and therefore will be omitted. The larger temperature fluctuations also prohibits reaching the force convergence criterion set in the LLG calculation which means that the LLG simulation runs until the maximal simulation time of $100\,\mathrm{ps}$ is reached. During this simulation time, skyrmions can spontaneously form and disappear which results in a finite topological charge measured at the end of the run.
In the future the real time dynamics of skyrmion creation and collapse may be the focus of the investigation. However, this approach may become unfeasible for situations where the skyrmion lifetime is very long compared to the typical time step in LLG calculations.
Finally, we highlight that with increasing temperature fluctuations we also find a larger variance in the number of skyrmions when averaging over the 5 different starting configuration for each pair $(T, B_z)$. This supports our interpretation that skyrmions are spontaneously created and annihilated by temperature fluctuations.

\subsection{Curie temperature using Monte Carlo}

\begin{figure}[htb]
    \centering
    \includegraphics[width=0.49\linewidth]{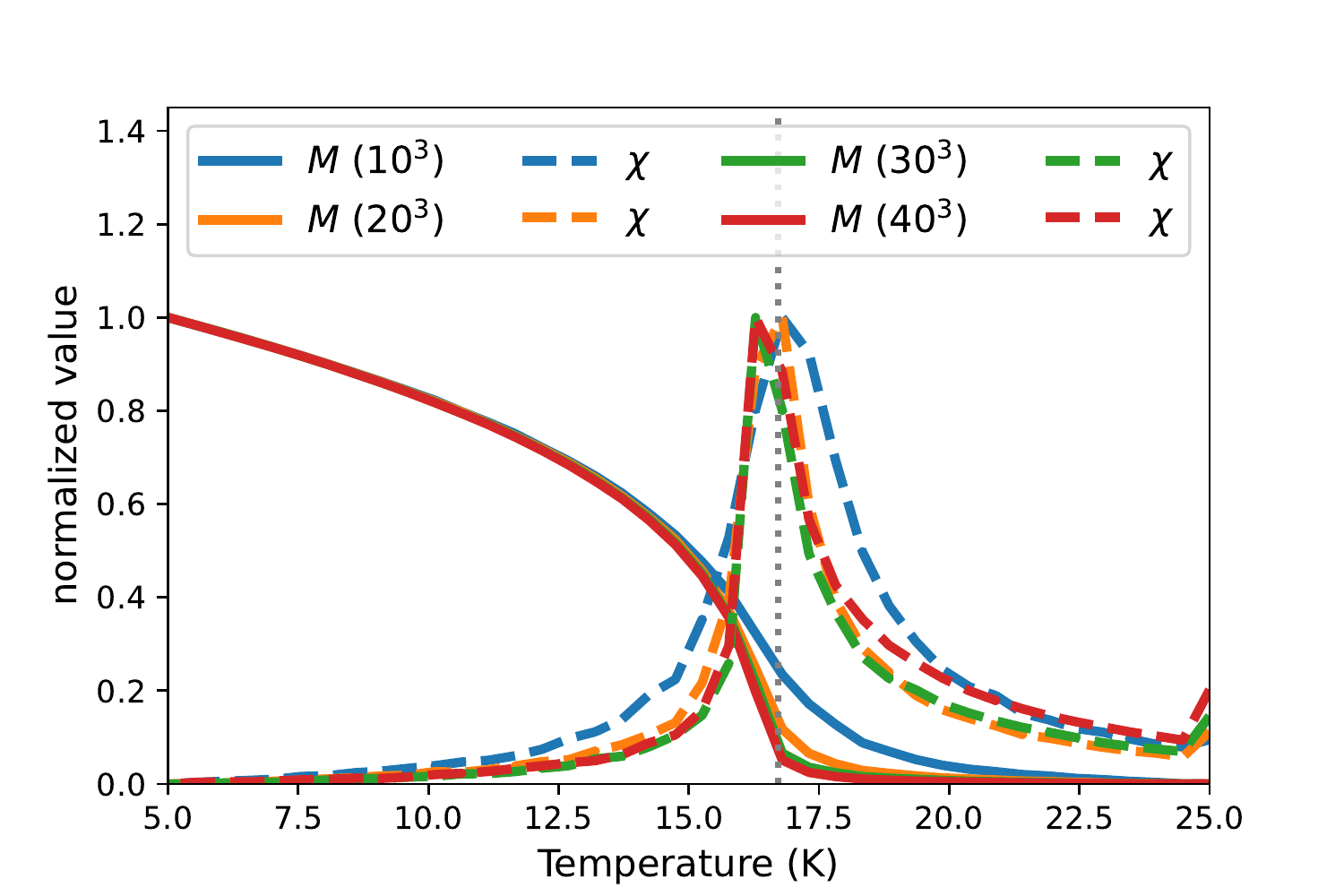}
    \caption{Results of the Monte Carlo calculations for a simple-cubic ferromagnetic with nearest neighbor $J_1=1\,\mathrm{meV}$ exchange interactions. Shown are results for $10\times10\times10$ to $40\times40\times40$ supercells where the solid lines show the normalized value of the total magnetization $M$ and the dashed lines the corresponding susceptibility $\chi$. The dashed vertical line indicate the expected value of the critical temperature at $T_c = 16.71\,\mathrm{K}$.}
    \label{fig:spiritMC}
\end{figure}

The Monte Carlo (MC) method is a well established tool in physics which, when applied to spin systems, allows to estimate the critical temperature of the magnetic ordering (Curie temperature) \cite{binderMC}. The \code{Spirit} code \cite{spirit-paper} implements a Metropolis algorithm which can be used from AiiDA-Spirit by choosing the \code{mc} simulation method (instead of the previously used \code{llg} method). We demonstrate the MC at the example of a simple-cubic ferromagnet with only nearest neighbor interactions $J_1=1\,\mathrm{meV}$. We perform calculations for varying supercell sizes between $10\times10\times10$ and $40\times40\times40$ with the MC parameters given in Table~\ref{tab:MCparam}. 
The results of the calculation are shown in Figure~\ref{fig:spiritMC} where, together with the total magnetization $M$, the isothermal susceptibility
\begin{equation}
    \chi = \frac{1}{k_\mathrm{B}T} ( \langle M^2 \rangle - \langle M \rangle^2 )
\end{equation}
with $M = |\frac{1}{N}\sum_i \vec{S}_i|$
the average magnetization of the sample is shown. We see that $M(T)$ converges with increasing supercell size indicating that boundary effects become less important. The corresponding susceptibilities show a diverging behavior at $T_c$. Our calculation results agree well with the expected value of $T_c = 1.44J_1/k_\mathrm{B}=16.71\,\mathrm{K}$. 
We stress that these calculations require a series of steps consisting of, for example, thermalization and decorrelation steps at each temperature value in the scanning interval. Within AiiDA-Spirit this complexity is conveniently absorbed in the \code{SpiritCalculation} which greatly facilitates the application of MC calculations.

\begin{table}[htb]
    \centering
    \caption{Parameters for the MC calculations of the simple-cubic ferromagnet discussed in the text. Note that the chosen settings result in temperature steps of $0.5\,\mathrm{K}$.}
    \begin{tabular}{c|c|c}
Parameter & Value & Description \\\hline
\code{n\_thermalisation} & $5000$ & Number of thermalization steps before \code{n\_samples} are taken \\
\code{n\_samples} & $250000$ & Number of samples taken in metropolis algorithm \\
\code{n\_decorrelation} & $2$ & Number of decorrelation steps \\
\code{n\_temperatures} & $40$ &  Number of temperature steps\\
\code{T\_start} & $25$ & Start of temperature scanning range \\
\code{T\_end} & $5$ & End of temperature scanning range \\
    \end{tabular}
    \label{tab:MCparam}
\end{table}

\subsection{Multi-scale modelling: $\gamma$--Fe}

We now demonstrate how the integration of the \code{Spirit} code into the AiiDA framework through the AiiDA-Spirit plugin can facilitate multi-scale modelling for magnetic materials. In this example we first calculate the exchange interaction parameters for $\gamma$--Fe using density functional theory which are then passed to the AiiDA-Spirit plugin for LLG simulations.

The $\gamma$ phase of Fe is a metastable high-temperature phase where the atoms crystallize in the fcc lattice \cite{Knoepfle2000, Sjoestedt2002}. This has a drastic consequence on the exchange interactions where, in contrast to the ferromagnetic bcc Fe, frustrated exchange interactions can lead to the formation of spin-spirals. Experimentally this structure of Fe can be realized in a Cu matrix \cite{Tsunoda_1989, TSUNODA1993133}. It is known that a variation of the lattice constant of $\gamma$--Fe can have drastic consequences for the magnetic ordering \cite{Sjoestedt2002}. Here, we investigate bulk crystals of $\gamma$--Fe for varying lattice constants between $a_\mathrm{lat}=3.2\,\mathrm{\AA}$ and $a_\mathrm{lat}=4.0\,\mathrm{\AA}$ around the lattice constant of Cu ($a_\mathrm{lat}=3.6\,\mathrm{\AA}$). 

\begin{figure}[htb]
    \centering
    \includegraphics[width=1.0\linewidth]{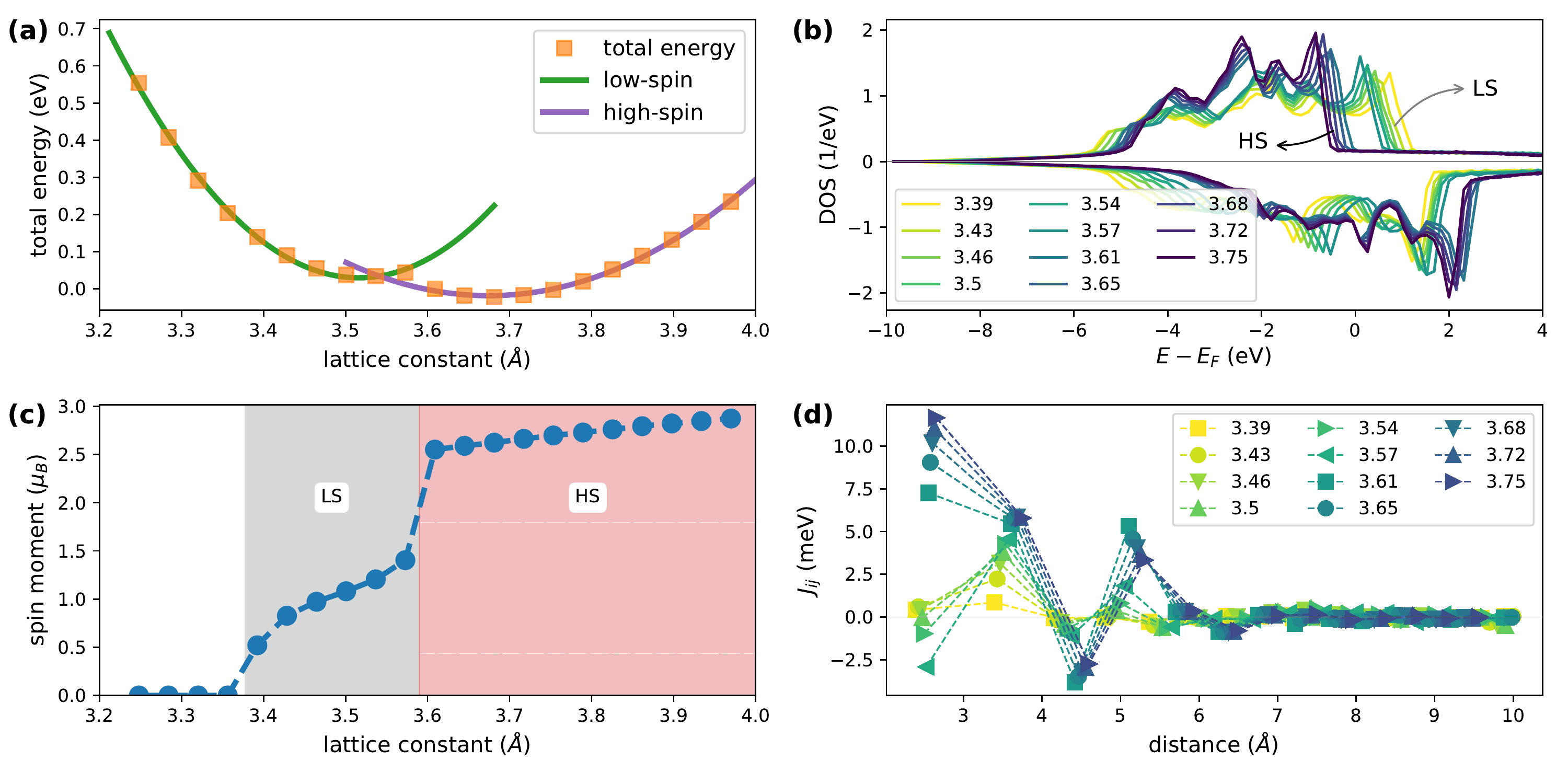}
    \caption{
    Results of the DFT calculations for $\gamma$--Fe. Total energy as a function of the lattice constant (a) where the green and violet lines show parabolic fits to low-spin ($a_\mathrm{lat}<3.6\mathrm{\AA}$) and high-spin ($a_\mathrm{lat} \ge 3.6\mathrm{\AA}$) states. (b) Corresponding density of states (value in the legend indicates the lattice constant) which clearly separates into low-spin (LS) with smaller exchange splitting and high-spin (HS) states with larger exchange splitting. Note that the positive (negative) values of the DOS indicate the majority (minority) spin channels. (c) Spin moment as a function of the lattice constant where the transition from non-magnetic (first four data points) over the low-spin state to the high-spin state is evident. (d) Resulting exchange coupling constants ($J_{ij}$) as a function of the pairwise distance between the Fe atoms in $\gamma$--Fe (the lattice constant is given in the legend).}
    \label{fig:DFT}
\end{figure}

Figure~\ref{fig:DFT} summarizes the results of the DFT calculations that were done with the AiiDA-KKR plugin (see methods section for numerical details). The total energy as a function of the lattice constant (shown in panel (a)) reveals a phase transition from the low-spin state (for $a_\mathrm{lat}<3.6\,\mathrm{\AA}$) to the high-spin state ($a_\mathrm{lat} \ge 3.6\,\mathrm{\AA}$) of $\gamma$--Fe as seen in the jump of the spin moment to $\mu_s>2.5\,\mu_B$. This coincides with a smaller exchange splitting seen in the density of states (panel (b)) and consequently a smaller value of the spin moment (c). For lattice constants below $3.37\,\mathrm{\AA}$ we find that the magnetic moment vanishes. Panel (d) shows the calculated exchange interactions $J_{ij}$ as a function of distance between two Fe atoms. Clearly, the sign of the nearest neighbor interaction shows the most drastic change with the transition from high-spin to low-spin state at smaller lattice constant of $\gamma$--Fe. While in the high-spin state the first and second nearest neighbor interaction are both ferromagnetic ($J_{ij}>0$), for the low-spin state the nearest neighbor interaction changes from being weakly ferromagnetic to antiferromagnetic ($J_{ij}<0$).

In the following, the consequences of this change for the magnetic ordering are investigated based on a series of LLG calculations using the AiiDA-Spirit plugin. In the DFT calculation we use the primitive cell which contains a single atom in the unit cell. For the spirit calculations we map the calculated exchange interactions onto the conventional unit cell consisting of 4 atoms. The parameters of the LLG simulations are summarized in Table~\ref{tab:LLGparam2}. We study the magnetic ordering in a $40\times40\times40\times4=256,000$ spins supercell without external magnetic fields and at temperature $T=0\,\mathrm{K}$. We choose open boundary conditions in order to not bias the eventually resulting spin-spiral wavelength by the periodicity of the supercell.

\begin{table}[htb]
    \centering
    \caption{Parameters for the LLG calculations for $\gamma$--Fe. Note that the simulation cell consists of $40\times40\times40\times4=256,000$ atoms due to the choice of the conventional unit cell with 4 atoms. The spin moment $\mu$ is extracted from the DFT calculation at the respective lattice constant and open boundary conditions are chosen. Parameters not listed here are set to the same value as in Table~\ref{tab:LLGparam}.}
    \begin{tabular}{c|c|c}
Parameter & Value & Description \\\hline
\verb!n_basis_cells! & $[40, 40, 40]$ & Size of the simulation cell \\
\verb!boundary_conditions! & [False, False, False] & Open boundary conditions \\
\verb!llg_temperature!         & $0$   &  Temperature [K] \\
\verb!external_field_magnitude! & $0$ &  Magnitude of the external field \\
\verb!mu_s! & $[\mu, \mu, \mu, \mu]$ & Spin moment [$\mu_B$]
    \end{tabular}
    \label{tab:LLGparam2}
\end{table}

\begin{figure}[htb]
    \centering
    \includegraphics[width=1.0\linewidth]{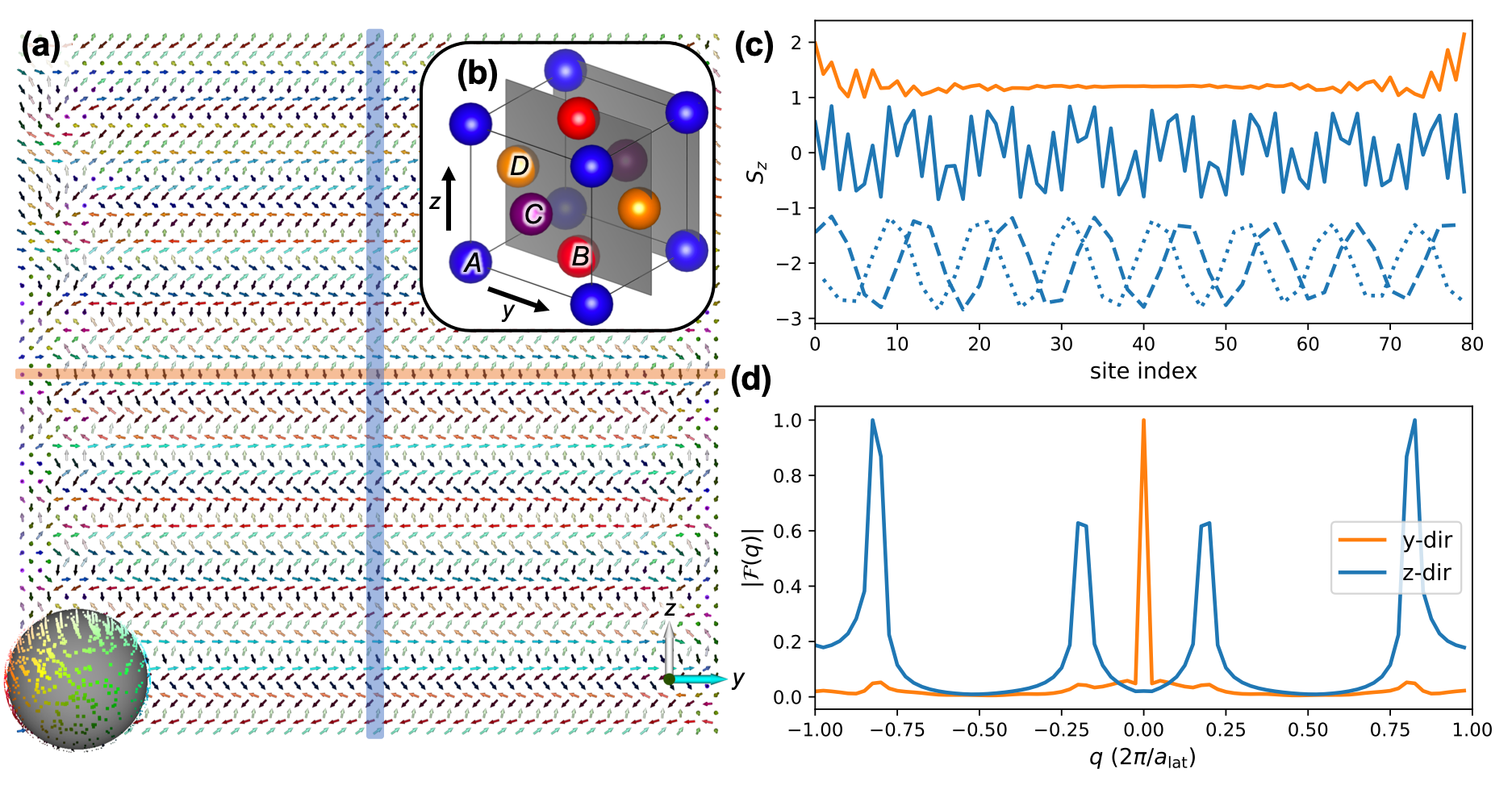}
    \caption{Spin-spiral ground state for $\gamma$--Fe at a lattice constant of $a_\mathrm{lat}=3.61\,\mathrm{\AA}$. \textbf{(a)} Cut through the central $yz$-plane of the $40\times40\times40$ supercell. Each unit cell consists of 4 Fe atoms on the sub-lattices \textit{A}, \textit{B}, \textit{C} and \textit{D} shown in the inset \textbf{(b)}. \textbf{(c)} $z$-component of the spin along the $y$- (orange line) and $z$-directions (solid blue line) indicated in \textbf{(a)} with the corresponding colored stripes. Each direction combines the spins from two adjacent $yz$-planes to cover more than a single sub-lattice per direction (i.e.\ sites in $y$-direction consist of atoms \textit{A}-\textit{B}-\textit{A}-\dots). The dashed and dotted lines in \textbf{(c)} are a decomposition of $S_z(z)$ into the sub-lattices \textit{A} and \textit{D}. Note that the orange and dashed and dotted blue lines are shifted by $\pm2$ with respect to the solid blue line. \textbf{(d)} Magnitude of the Fourier transformation of $S_z$ along the $y$- and $z$-directions, i.e.\ solid orange and blue lines in \textbf{(c)}, respectively.}
    \label{fig:SpiritGammaFe0}
\end{figure}

We start the discussion of the LLG calculations with the results for $\gamma$--Fe in the lattice constant of Cu ($a_\mathrm{lat}=3.61\,\mathrm{\AA}$). Figure~\ref{fig:SpiritGammaFe0}(a) shows the resulting spin texture at the end of the LLG calculation for the central layer of spins in the $yz$-plane (shown in the inset (b)). We can see that a spin-spiral forms in $z$-direction with ferromagnetically ordered spins in $y$-direction. At the open boundaries  we see that the missing neighbors on one side influence the magnetic ordering that deviates from the spin-spiral in the center for a distance of about 5 lattice constants. In order to quantify the spin-spiral we pick the two cardinal directions in this plane (indicated by blue and orange lines in (a)) and extract the $z$-component of the spin $S_z$. We combine the projections onto the $yz$-plane from two adjacent layers of spins (indicated by the two grey planes in (b)) to not restrict our analysis to a single sub-lattice only. This allows to describe also antiferromagnetic structures in the sub-lattices with ferromagnetic ordering within one sub-lattice, which will be important later. Figure~\ref{fig:SpiritGammaFe0}(c) shows that, except for boundary effects, $S_z$ stays constants when following the $y$-direction. Along the $z$-axis we see a complex beating pattern with the site index that can be decomposed into two $\pi/2$-shifted spin-spirals in the two different sub-lattices. The corresponding Fourier transformation
\begin{equation}
    \mathcal{F}_j(q) =\frac{1}{\sqrt{2\pi}}\int e^{-i q r_j} S_z(r_j) \, \mathrm{d}r_j
    \label{eq:fft1}
\end{equation}
with $j = y, z$ computed with the fast Fourier transform algorithm (FFT) is shown in (d). As expected, the FFT of the predominantly ferromagnetically ordered spins along the $y$-direction $|\mathcal{F}_y|$ mainly shows a signal at $q=0$ whereas $|\mathcal{F}_z|$ shows the appearance of four peaks at $q\approx\pm0.2\,\frac{2\pi}{a_\mathrm{lat}}$ and $q\approx\pm0.8\,\frac{2\pi}{a_\mathrm{lat}}=(1-0.2)\,2\pi / a_\mathrm{lat}$ which are attributed to the two $\pi/2$-shifted oscillations in the two sub-lattices. We point out that the reflection symmetry around $q=0$ is a consequence of the real-valued input to the FFT and is therefore not discussed further.
As seen from the FFT in $y$-direction and from the corresponding spin texture in (a) the spins in the direction perpendicular to the propagation direction of the spin spiral (i.e.\ the $z$-direction in this example) are ordered ferromagnetically. Therefore the spin-spiral wavevector is $\vec{q} = (0,0, 0.2)\,2\pi/a_\mathrm{lat}$.
After having characterized the spin-spiral ground state of $\gamma$--Fe we continue with a discussion of the magnetic ordering depending on the changing exchange coupling parameters with changing lattice constant.

\begin{figure}[htb]
    \centering
    \includegraphics[width=1.0\linewidth]{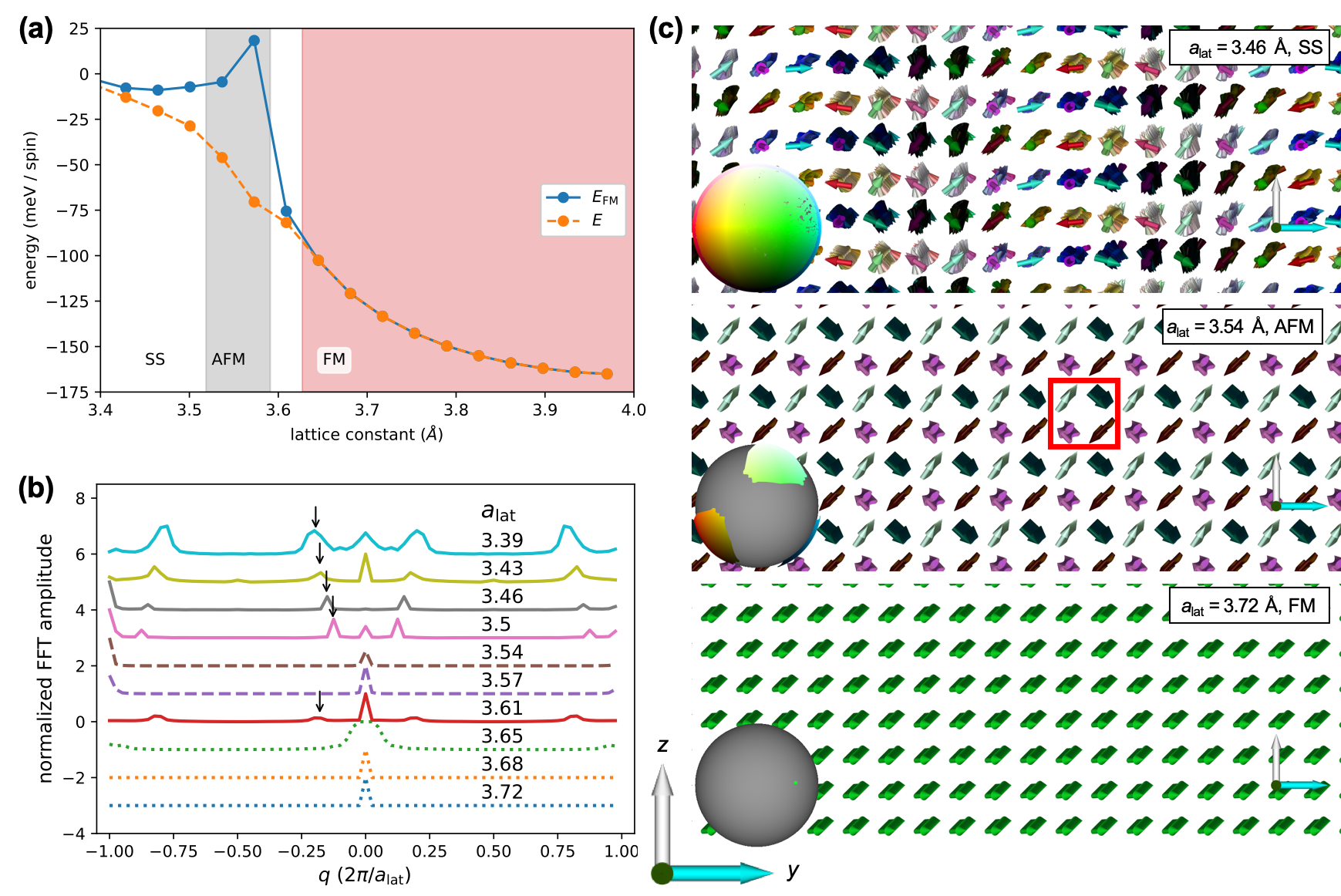}
    \caption{Magnetic ground state in $\gamma$--Fe from spin-dynamics simulations via the AiiDA-Spirit plugin. (a) Calculated energies per spin of the final state after an LLG calculation. Each data point uses the exchange constants computed from DFT (cf.\ Fig.~\ref{fig:DFT}d). The solid blue and dashed orange lines indicate the energy computed starting from random spin configuration or the ferromagentic (FM) state. The white, grey and red shaded areas indicate if,respectively, a spin-spiral (SS), antiferromagnetic (AFM) or FM ordering is found to be the ground state. (b) Normalized Fourier transform of the $z$-component of the magnetization in the $yz$-plane for different lattice constants, shifted for clarity. Solid lines correspond to SS, dashed lines to AFM and dotted lines to FM solutions, repectively. The arrows highlight the principal wavenumber of the spin-spiral and highlight their change with the lattice constant. (c) Visualization of representative spin structures for (from top to bottom) SS, AFM and FM states where the red box in the AFM structure highlights a unit cell with the 4 sub-lattices.}
    \label{fig:SpiritGammaFe}
\end{figure}

Figure~\ref{fig:SpiritGammaFe} summarizes the LLG calculations for $\gamma$--Fe for varying lattice constants using the respective set of exchange parameters shown in Fig.~\ref{fig:DFT}(d). The lines in Figure~\ref{fig:SpiritGammaFe}(a) show the energy at the end of the LLG calculation starting either from a random spin configuration ($E$, dashed orange) or from the ferromagnetic ($E_\mathrm{FM}$, solid blue) state. We find that for lattice constants $a_\mathrm{lat} \ge 3.65\,\mathrm{\AA}$ the ferromagnetic state minimizes the energy ($E-E_\mathrm{FM}=0$). We attribute this to the increasing ferromagnetic interaction for nearest neighbor spins in the high-spin state which was discussed with Figure~\ref{fig:DFT}(d). At smaller lattice constants ($a_\mathrm{lat} \leq 3.6\,\mathrm{\AA}$), the ferromagnet (FM) is not the ground state anymore. Here we find either a spin-spiral (SS) ground state or an antiferromagnetic (AFM) phase. Figure~\ref{fig:SpiritGammaFe}(c) shows three representative images of the SS, AFM and FM states.

In the SS state the magnetization rotates from left to right (i.e.\ along the $y$-axis) and shows antiparallel alignment of the rows in $z$-direction. Along the $x$-axis (direction perpendicular to the drawn plane) the spins are aligned ferromagnetically, except for boundary effects at the open ends of the simulation cell (seen in the direction of the first layer of spins). 
In $z$-direction, adjacent layers are antiferromagnetically ordered. Thus the spin-spiral wavevector for these lattice constants has the form $\vec{q}=(0, q, 1)\,2\pi/a_\mathrm{lat}$ which due to the cubic symmetry of the crystal is equivalent to $\vec{q}=(q, 0, 1)\,2\pi/a_\mathrm{lat}$. Note that $\vec{q}_\mathrm{AF} = (0,0,1)\,2\pi/a_\mathrm{lat}$ is the antiferromagnet because the distance between two layers in the $(0,0,1)$ direction of the fcc lattice is $a_\mathrm{lat}/2$. 

In the AFM phase the direction of the spins separate into four sub-lattices that correspond to the 4 atoms in the conventional fcc unit cell. Within each sub-lattice the spins are aligned parallel and form a right angle with their neighboring spins from different sub-lattices. This is highlighted with a red box in the middle panel of (c). In the FM phase (lower panel) all spins point in the same direction. Note that in all these calculations the spins can collectively rotate since we neglected contributions from single-ion anisotropies and do not apply an external field.

The summed magnitude of the Fourier transform along the three cardinal axes
\begin{equation}
    |\mathcal{F}(q)|=\sum_{j=x,y,z}|\mathcal{F}_j (q)|
    \label{eq:fft2}
\end{equation}
is shown in Figure~\ref{fig:SpiritGammaFe}(b) for different lattice constants. Note that we have summed here over the symmetry-equivalent directions along the $x$- $y$- and $z$-directions because of the rotational invariance of the complete spin-structure. Starting from the smallest lattice constant of $a_\mathrm{lat}=3.39\,\mathrm{\AA}$ we see a peak in $|\mathcal{F}(q)|$ at $\vec{q} = (\pm 0.195,0,1)\,2\pi/a_\mathrm{lat}$ which corresponds to a wavelength of the spin-spiral of $5.13\,a_\mathrm{lat}$. Here we focus our discussion on the peak at smaller $q$ values as the same arguments hold for the second peak at $2\pi/a_\mathrm{lat}-q$ as discussed above. With increasing lattice constant the spin-spiral wavelength increases to $7.82\,a_\mathrm{lat}/2$ ($\vec{q}=(\pm 0.128,0,1)\,2\pi/a_\mathrm{lat}$) at a lattice constant of $a_\mathrm{lat}=3.5\,\mathrm{\AA}$. For $3.5\,\mathrm{\AA}<a_\mathrm{lat}<3.6\,\mathrm{\AA}$ the AFM state is found, which in the Fourier transform is characterized by the dominating peak at $q=\pm 2\pi/a_\mathrm{lat}$. Note that we still get a considerable signal at $q=0$ because we sum over all three cardinal directions and there is ferromagnetic ordering along one direction (see Fig.~\ref{fig:SpiritGammaFe}(c)). As discussed above, for larger lattice constants the spin-spiral state briefly shows up again at $a_\mathrm{lat}=3.61\,\mathrm{\AA}$ with a wavelength of $5.59\,a_\mathrm{lat}$ which however has ferromagnetically ordered spins in both directions perpendicular to the direction of spin-spiral propagation ($\vec{q}=(\pm 0.179,0,0) \,2\pi/a_\mathrm{lat}$) until for $a_\mathrm{lat}\ge3.65\,\mathrm{\AA}$ the FM state is found which only shows a significant Fourier amplitude at $q=0$.

\begin{figure}[htb]
    \centering
    \includegraphics[width=1.0\linewidth]{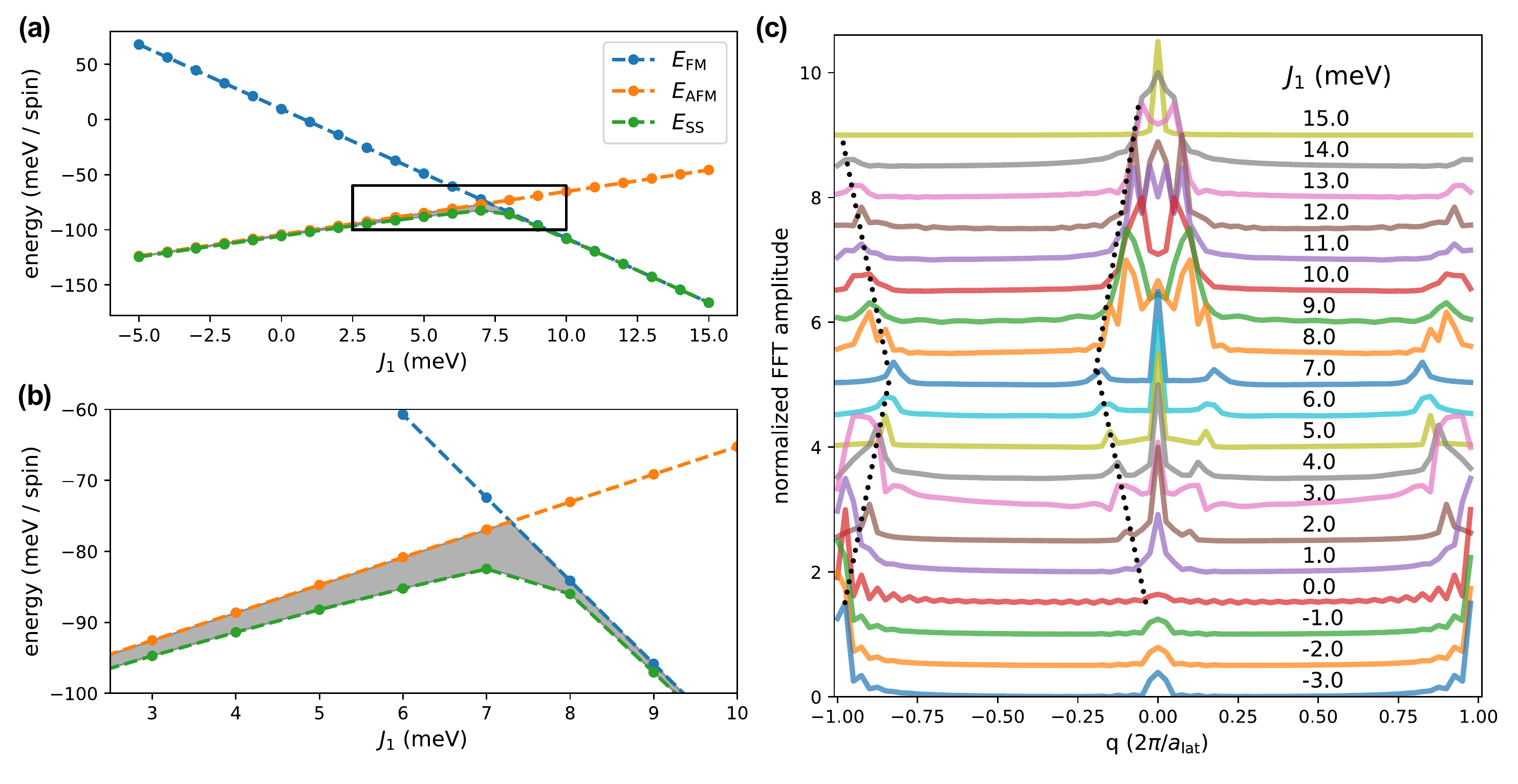}
    \caption{(a) Spin-spiral (SS) energies as a function of the nearest neighbor interaction $J_1$ in comparison to ferromagnetic (FM) and antiferromagnetic (AFM) states. Exchange coupling constants were taken from the $a_\mathrm{lat}=3.61\,\mathrm{\AA}$ calculation and the nearest neighbor couplings $J_1$ were modified in the range from $-5$ to $15\,\mathrm{meV}$. (b) Zoom into the transition region marked by the black box in (a). The grey area highlights the energy gain in the SS state compared to FM or AFM states. (c) Normalized Fourier transform of the $z$-component of the magnetization in the $yz$-plane. The spectra are shifted for clarity. Dashed lines indicate where $E_\mathrm{AFM} - E_\mathrm{SS} < 2\,\mathrm{meV}$ and dotted lines are used for $E_\mathrm{FM} - E_\mathrm{SS} < 2\,\mathrm{meV}$. The solid and dashed grey lines are guides to the eye highlighting the change in the spin-spiral wavevector with $J_1$.}
    \label{fig:spiritFFTscaledJ0}
\end{figure}

The appearance of the AFM phase for $3.5\,\mathrm{\AA}<a_\mathrm{lat}<3.6\,\mathrm{\AA}$ can be attributed to the sign change of the nearest neighbor interaction from ferro- to antiferromagnetic. To verify this hypothesis we employ a series of LLG calculations through AiiDA-Spirit. We chose to start from the setup of the calculation for $a_\mathrm{lat}=3.61\,\mathrm{\AA}$, which was found to reproduce the spin-spiral phase. We then modify the strength of the nearest neighbor interaction $J_1$ ranging from $-5\,\mathrm{meV}$ to $+15\,\mathrm{meV}$ and run LLG calculations starting from the FM state, from the AFM state, from the SS phase and random spin orientations. For the AFM state we construct the row-wise AFM orientation of the spins. All other parameters for the LLG calculation are kept constant. In total this is another set of 84 \code{SpiritCalculations} where we find that starting from a random spin configuration coincides with starting from the SS. The random starting point is therefore omitted in the following discussion. Figure~\ref{fig:spiritFFTscaledJ0} (a,b) show the dependence of the energy per spin at the end of the LLG calculations. We point out that the LLG calculations for the FM and AFM states are converged in the very first iteration which indicates that the FM and AFM phases are local minima in the energy landscape. We find that the SS state is lowest in energy with a maximal energy gain of $\sim7\,\mathrm{meV/spin}$ in the transition region where $E_\mathrm{FM}-E_\mathrm{AFM}$ changes sign.
From the final spin structure of the spin-spiral solution we proceed with an analysis of the Fourier components as introduced in equations \ref{eq:fft1} and \ref{eq:fft2}. This is shown in Figure~\ref{fig:spiritFFTscaledJ0}(c). As highlighted by the grey line, we see an increase in the spin-spiral wavevector with increasing $J_1$ up to the point where $E_\mathrm{FM}$ and $E_\mathrm{AFM}$ cross around $J_1 = 7.3\,\mathrm{meV}$. As in the previous analysis for changing lattice constant we find that the spin-spiral state is characterized by two wavevectors at $q$ and $2\pi/a_\mathrm{lat}-q$ (grey dashed line). Furthermore, the Fourier transform for all states show significant signals at $q=0$ (indicating parallel spins) and $q=\pm2\pi/a_\mathrm{lat}$ (indicating antiparallel spins).

Overall we can conclude that the resulting spin-texture in the 256,000 spin unit cell with open boundary conditions is a result of the complex competition of distance-dependent exchange couplings that favor ferromagnetic or antiferromagnetic alignments of spins or can compete and stabilize spin-spiral ground states.

\section{Discussion}

In this paper we have presented the AiiDA-Spirit plugin that connects the spin-dynamics code \code{Spirit} to the AiiDA framework. AiiDA enables high-throughput calculations while automatically keeping track of the data provenance \cite{aiida1}. We have demonstrated the capabilities of the AiiDA-Spirit plugin with three examples; (i) high-throughput spin-dynamics calculations based on the Landau-Lifshitz-Gilbert (LLG) equation for a toy model that shows skyrmions, (ii) Monte Carlo calculations for finding the critical temperature of a simple-cubic model ferromagnet, and (iii) multi-scale modelling combining density functional calculations with spin-dynamics simulations for $\gamma$--Fe.

In our high-throughput LLG calculations we performed more than 6000 simulations of a model system consisting of a 2D lattice of spins in the simple-cubic lattice. The model parameters were chosen such that topologically nontrivial skyrmions appear in the magnetic textures. We varied the temperature and the external magnetic field as external parameters and investigate the change in the topological charge, which is a measure of the number of skyrmions that appear in the system.
We find that, starting from $T=0$, the transition to the homogeneous ferromagnetic phase happens at lower magnetic fields. At a certain critical temperature however the number of skyrmions starts increasing again. We interpret this as the surpassing of the energy barrier for skyrmion formation which can be overcome by temperature fluctuations of the spins. This goes hand in hand with a larger variance in the topological charge that we measure from averaging multiple runs for each pair of $(T, B_z)$. These calculations demonstrate the possibility to employ the AiiDA-Spirit plugin for high-throughput spin-dynamics simulations which make parameter exploration easier accessible. 

In our Monte Carlo calculations we showed how the complex series of calculations necessary for finding the ordering temperature of a simple-cubic ferromagnet (several calculations across the transition region from ferromagnetically ordered to paramagnetic state have to be performed) can be found from a single \code{SpiritCalculation} of the AiiDA-Spirit plugin. Our simulation result is in good agreement with the theoretically expected result. The ease-of-use for these calculations facilitate the incorporation of AiiDA-Spirit calculations in complex workflows in materials informatics for magnetic materials. Here, finding the critical temperature of a magnetic material is a very common problem. 

Finally, we discussed the use case of LLG calculations for the study of the magnetic ordering of $\gamma$--Fe, which is the high-temperature fcc phase of Fe. From experiments, where Fe clusters were embedded in a Cu matrix, it is known that a spin-spiral ground state with wavevector $\vec{q}=(0.1, 0, 1)\,2\pi/a_\mathrm{lat}$ is found for $\gamma$--Fe around the lattice constant of Cu \cite{Tsunoda_1989, TSUNODA1993133}. Note that $q=(0,0,2\pi/a_\mathrm{lat})$ correspond to the antiferromagnet since the distance between two layers in the $(0,0,1)$ direction of the fcc lattice is $a_\mathrm{lat}/2$. Theoretically, this wavevector was reproduced from first-principles calculations with good agreement where $\vec{q} = (0.15, 0, 1) \,2\pi/a_\mathrm{lat}$ \cite{Knoepfle2000} and $\vec{q} = (0.16, 0, 1) \,2\pi/a_\mathrm{lat}$ \cite{Sjoestedt2002} were found. However, a significantly different lattice constant compared to the lattice constant of Cu is required for $\gamma$--Fe in the calculation compared to the experiments, which makes this agreement unsatisfactory \cite{Sjoestedt2002}.
In our work, instead of looking for the spin-spiral energies from first-principles calculations, we chose to explore the predictive power of a combination of DFT and LLG calculations for $\gamma$--Fe with changing lattice constants. We found a change from the low-spin to high-spin ground state in our DFT results that were performed with the \code{JuKKR} code \cite{jukkr} through the AiiDA-KKR plugin \cite{aiida-kkr-paper, aiida-kkr-code}. This agrees well with earlier DFT calculations where a similar change in the spin moment of the Fe atoms from $\mu\sim1\,\mu_\mathrm{B}$ to $>2.5\,\mu_\mathrm{B}$ is seen \cite{Knoepfle2000, Sjoestedt2002}. 

In contrast to the spin-spiral energy calculations of Refs.~\cite{Knoepfle2000, Sjoestedt2002} we calculate the exchange parameters for the extended Heisenberg model from the method of infinitesimal rotations \cite{Liechtenstein1987} around the collinear, ferromagnetically ordered state. These parameters are then used in the \code{SpiritCalculations} where the collective magnetic ordering is investigated in a 256,000 spin supercell. We find a strong influence of exchange interactions on the lattice constant of $\gamma$--Fe which results in a competition of ferromagnetic, antiferromagnetic and spin-spiral orderings. In our analysis of the spin-spiral wavevectors we chose to study the Fourier transform of the $z$-component of the spin around the three cardinal axes which are symmetry-equivalent in our approach. 
At the lattice constant of Cu ($a_\mathrm{lat}=3.6\,\mathrm{\AA}$) we find a spin-spiral with $\vec{q}=(0.2, 0, 0)\,2\pi/a_\mathrm{lat}$ in contrast to the spin-spiral $\vec{q}_\mathrm{exp} = (0.1, 0, 1)\,2\pi/a_\mathrm{lat}$ found experimentally which has an antiferromagnetic component \cite{Tsunoda_1989}. 
We attribute this discrepancy to neglecting the change in the spin moment with the spin-spiral wavevector in our simulations based on the Heisenberg Hamiltonian. This change is known to be significant and can be as large as $0.8\,\mu_\mathrm{B}$ \cite{Sjoestedt2002}.
For lattice constants $a_\mathrm{lat} \le 3.5$ we do find the correct spin-spiral with $\vec{q}=(q,0,1)\,2\pi/a_\mathrm{lat}$ that reappears after the spin-spiral with ferromagnetic ordering perpendicular to $\vec{q}$ transforms into the ordered antiferromagnet. With smaller lattice constant the spin-spiral wavevector increases from $\vec{q} = (0.13, 0, 1)\,2\pi/a_\mathrm{lat}$ at $a_\mathrm{lat}=3.5\,\mathrm{\AA}$ up to $\vec{q} = (0.2, 0, 1)\,2\pi/a_\mathrm{lat}$ at $a_\mathrm{lat}=3.39\,\mathrm{\AA}$ which is in reasonable agreement with earlier calculation results \cite{Knoepfle2000, Sjoestedt2002}.
Incorporating a change of the spin moment with the spin-spiral wavevector might further improve our agreement to the earlier \textit{ab initio} results of Kn\"opfle \textit{et al.} \cite{Knoepfle2000} and Sj\"ostedt and Nordstr\"om \cite{Sjoestedt2002} and also the experimental spin-spiral wavevector \cite{Tsunoda_1989}.

The change of the ordering to the antiferromagnetic state and then the reappearance of the spin-spiral state at even smaller lattice constant compared to the lattice constant of Cu on the other hand agrees well with the previously stated observation of competing magnetic orders, which are very close in energy and could coexist \cite{Sjoestedt2002}.
We further demonstrated the sensitivity of the magnetic ordering with a numerical experiment where we chose to modify the strength of the nearest neighbor exchange interaction $J_1$. The resulting strong change in the spin-spiral wavevector and the magnetic ordering highlights the rich energy landscape that is underlying the complex magnetic ordering in $\gamma$--Fe.

In conclusion, we have shown how augmenting spin-dynamics calculations with the \code{Spirit} code through the AiiDA-Spirit plugin enables high-throughput spin-dynamics simulations via the AiiDA infrastructure. This was applied to model systems and, in combination with DFT calculations through the AiiDA-KKR plugin, to the multi-scale problem of the magnetic ordering in $\gamma$--Fe. Our results demonstrate that typical spin-dynamics simulations benefit from the possibility to run a large number of calculations in a high-throughput fashion. Automation of \code{SpiritCalculations} through AiiDA can be a great asset when complex model parameter spaces (i.e.\ external fields, temperatures, different geometries, \dots) are screened in order to find structure-property relations of magnetic materials. The feature of AiiDA to keep track of the data provenance is here indispensable to get reproducible results and to eventually engineer recipes for the creation and control of unconventional topological solitons in magnetic structures such as skyrmions or hopfions in the future.

\section*{Conflict of Interest Statement}

The authors declare that the research was conducted in the absence of any commercial or financial relationships that could be construed as a potential conflict of interest.

\section*{Author Contributions}

PR and JRB programmed the first version of AiiDA-Spirit and MS and FR contributed to the further development of the plugin where FR was responsible for the \code{spin\_view} functionality of AiiDA-Spirit.
PR performed the DFT calculations and PR and MS performed the AiiDA-Spirit calculations. 
All authors discussed the results and contributed in writing the manuscript.

\section*{Acknowledgments}

We acknowledge support by the Joint Lab Virtual Materials Design (JL-VMD) and thank for computing time granted by the JARA Vergabegremium (project number jara0191) and provided on the JARA Partition part of the supercomputer CLAIX at RWTH Aachen University. This work was funded by the Deutsche Forschungsgemeinschaft (DFG, German Research Foundation) under Germany's Excellence Strategy – Cluster of Excellence Matter and Light for Quantum Computing (ML4Q) EXC 2004/1 – 390534769.

\section*{Data Availability Statement}
The datasets generated for this study can be found in the materials cloud \cite{doi-dataset}.

\bibliography{references}

\end{document}